# Why lot: How sortition came to help representative democracy

*Maurizio Caserta[1], Alessandro Pluchino[2], Andrea Rapisarda [2],[3], Salvatore Spagano [1]*

[1]Dipartimento di Economia ed Impresa, Università di Catania, Italy
[2]Dipartimento di Fisica e Astronomia "Ettore Majorana" and INFN, Università di Catania, Italy
[3]Complexity Science Hub Vienna, Austria

**Abstract**
In this paper we discuss the problems of modern representative democracy and we look at the selection of legislators by lot as a way to tame some of the drawbacks of that system. It is recalled at the beginning that resorting to sortition for the selection of public officers used to be a popular way of taming factionalism in public affairs. Factionalism is assumed to be detrimental to public affairs as public officers may favour their own faction (a tribe or a party) instead of pursuing the general interest. Moreover they tend to overinvest in strengthening their power, thus engaging in power struggles with opposing factions, unlikely to benefit society. In this respect we present a new mathematical model aiming at describing a more efficient parliament where sortition is brought to bear. It will be shown that starting from a parliament working with two parties (or coalitions), where the costs of representative democracy are quite apparent through the detrimental effects of party discipline, one can beneficially move towards a parliament where independent legislators, randomly selected from the population of constituents, sit alongside elected members who belong to a party and are subjected to party discipline. The paper shows that increasing the number of independent legislators up to a point enhances the efficiency of the parliament and puts into check the factionalism likely to arise from party discipline.

# 1. Introduction

Resorting on random procedures when it comes to grant powers, assign public functions or take collective decisions has been interpreted in different ways in the recent literature.[1] Lockard (2003), for example, looks at those procedures as ways of contrasting rent-seeking behaviour. Similarly, Stone (2011) and Delannoi, Dowlen and Stone (2013) emphasize their role in fighting corruption. Others, like Callenbach and Phillips (2008) or Callenbach, Phillips and Sutherland (2008), look at sortition as a way to accurately represent a diverse population in a smaller subset. Finally Stone (2016) shows under what conditions sortition makes sure societies achieve allocative justice.

Sortition has also been considered as a way to fix some specific drawbacks of contemporary political systems. See, for example, the thought-provoking proposals in Frey (2017)[2].

Following the line set in Dowlen (2008 and 2017), this paper places sortition among the instruments designed to fight concentration of political power. Some of the undesirable effects of political power, especially those currently more widespread, like corruption, elitism, self-referential behaviour, are often attributed to such a concentration. Viewed in this way, sortition parallels other mechanisms designed to keep political power in check, like division of powers, universal suffrage, political term limits, rotational assignment.

As a matter of fact, through history, sortition has done quite a good job in harnessing a specific source of power concentration, i.e. factionalism. By factionalism we mean a tendency to strengthen bonds within a group, usually in opposition to other groups, with the purpose of gaining more power. When such bonds are particularly strong, factionalism may lead to excessive concentration of power. If a faction holds a government office, it may use the benefits arising from that office to strengthen those bonds even more, with the purpose of fighting opposing factions. Eventually this may lead to relax the pursuance of the general interest.[3]

Factionalism[4] is not a new development. Throughout history it has taken up various forms and has worked through various mechanisms, as Dowlen (2008) already argued. Its current parliamentary version is party discipline. Party discipline is called for whenever internal cohesion is required to stress group identity against opposing parties.[5] Resuming lot as a way to contrast factionalism, in this paper we present a mathematical model of a parliament with a bipolar party composition, where an additional component of legislators, selected by lot and independent of parties, is introduced. In line with a previous exploratory studies (Pluchino *et al.* 2010 and Pluchino *et al.*, 2011), the model shows the effectiveness of sortition in mitigating the detrimental effects of party discipline, and thus of inefficient public decision-making.

---

[1] Arrow (1963, pp. 20-21) argued that probabilistic methods could even overcome the "impossibility" of his own theory.

[2] See the related responses by Köppl-Turyna (2018) and Tridimas (2018).

[3] As any other tyranny, the *tyrannie de la majorité*, which De Tocqueville (1850) cautioned against, may be considered as a case of factionalism.

4 Chan and Man (2012) point to a number of incentives underlying factionalism that can facilitate its inception.

[5] Eguia (2011) identifies some incentives to make party discipline effective. For a broader overview of the topic, see Bowler, Farrell and Katz (1999).

The paper is organized as follows. The second section surveys how sortition has been used in history as a way to fight factionalism. The third section introduces the mathematical model and the fourth section addresses the analytical calculation of the parliament's efficiency in the two limiting cases of only two parties or only independent legislators. Section 5 presents the results for the most general case of a mixed parliament, with both parties and independent legislators, while section 6 looks for its maximum efficiency. Finally, section 7 discusses some interesting proposals, based on the paper's results, for repairing representative democracy and section 8 concludes the paper. Details of analytical calculations are presented in four appendices.

**2. Sortition versus factionalism**

The first instance of sortition as a way to organize power goes back to ancient Athens, as is very well known. However, we do not have unambiguous evidence of the time and the circumstances when all this started.[6] Unlike what happened during the Italian Renaissance, in ancient Athens sortition was not intended to extend participation; women, minors and slaves remained outside political life. Instead, sortition came to be seen as a way, among others, to contrast the tendency to establish solid bonds among individuals, lest such bonds could jeopardize the attainment of social welfare.

Therefore, sortition was not jus a way to select politicians. It must be said, in passing, that a large literature exists already on this particular view of the issue.[7] Indeed, sortition was intended also differently, as a way to mould and direct political action. One can look at the way constituencies were reformed in ancient Athens to find evidence of this other purpose of sortition. It was Kleisthenes who put ten tribes in place of the long-established four tribes.

According to Aristotle (1986)[8], Kleisthenes broke up the Athenian region into thirty groups called trittyes: ten from the city, ten from the coast and ten from the inland. Coming from the same region, each trittys was highly homogeneous. Therefore, the trittyes were very likely to turn into opposing factions, as they represented diverse interests. Turning to lot avoided this occurrence, as each tribe was made up of three different trittyes, one coming from the city, one from the coast and one from the inland. Establishing tribes not according to family ties but according to place of birth made clear the true purpose of lot: preventing the formation of strong bonds and emasculating the existing ones. In turn, those ten tribes would make up the population the 500 members of the Boule, established by Klisthenes in 508-507 B.C., would be taken from by lot. Fifty representatives would be selected from each tribe.

That sortition by lot in ancient Athens was mainly designed to fight factionalism is borne out by two accompanying features of sortition. First, public officers could not hold the same office twice. The only exception was the Boule. In that case the office could be held twice. This is explained by the limited number of eligible individuals with respect to the number of selected representatives. If only one term had been possible, it might have been difficult to fill all the available seats in the Boule. Second, office could be held only for one year. This time was too short for factionalism to re-emerge. Hence, lot, term limits, short time in office, all seem to work in the same direction: preventing personal or group bonds from becoming stable and strong. Failing all these devices, the rule of majority would very likely yield opposing blocks fiercely competing for power.

---

[6] On the non-political origins of sortition, see Dowlen (2017), p. 31 ff.
[7] Delannoi and Dowlen (2016) lists several old and current ways for implementing political sortition.
[8] P. 164.

Sortition became very popular in Italian city-states between the late Middle Ages and the Renaissance. The '*Brevia*' and the '*Scrutinio e tratta*'[9] were the two main instances. The '*Brevia*' spread across Northern Italy between the twelfth and the thirteenth century and survived in the version that got established in Venice until the beginning of the nineteenth century. The '*Scrutinio e tratta*' was developed in Florence at the beginning of the fourteenth century and survived for one and a half century.

There is evidence that in the 'Scrutinio' sortition was precisely designed to overcome opposition among existing factions. After the death of the Duke of Lucca who in 1328 gave back to the Florentine the right to choose their government, how to achieve that became an issue to be discussed. In that discussion it clearly emerged that the ultimate purpose of any reform should be getting rid of factions. [10] A few months later Florence adopted the '*Scrutinio e tratta*' scheme. It is interesting to note that, alongside that scheme, it was ruled out that the relatives of those who had been selected through sortition for a public office could be selected themselves for the same office. Again sortition proves to be designed mainly to stave off fights among opposing factions, the underlying assumption being that those fights are not good for society.

Similar evidence exists in the case of '*Brevia*'. The Statute of the city of Parma introduced the 'Brevia' in 1233; there, sortition is clearly presented as a means to avoid *contentiones*.[11] Interestingly, the '*Brevia*' and the '*Scrutinio*' mirrored each other. In the '*Brevia*' sortition came at the beginning of the whole selection procedure through the selection by lot of a pool of electors (nominators in Venice). Once selected the electors would go on to vote or nominate their representatives. The '*Scrutinio*' would work the other way round. First, the pool of representatives would be elected. Then, starting from that pool, public offices would be allocated by lot. Using sortition interchangeably, at different stages of the selection procedure, shows that the purpose of sortition is really fighting factionalism, just like ancient Greece.

Furthermore, it shows that random procedures can coexist with teleological ones. At that time, therefore, in the Italian city-states, running across the Renaissance, there was no presumption that random selection of public officers would be superior to teleological selection through voting. There was no presumption, that is, that wisdom is smoothly distributed across the population or that qualitative differences among individuals eligible to public office were irrelevant, with the effect of making rational, intentional and teleological selection entirely useless.

---

[9] Cfr. Najemy (1982) for '*scrutinio e tratta*' and Wolfson (1899) for '*brevia*'.
[10] "*Dappoich'è Fiorentini ebbono novelle della morte del duca, ebbono più consigli e ragionamenti e avvisi, come dovessono riformare la città di reggimento e signoria per modo comune, acciocché si levassono le sette tra' cittadini*": Villani (1845), III, p.103.
[11] "*Capitulum ad evitandum quod aliquis qui non sit de consilio generali debeat stare ad sortes recipiendas, et ad evitandum contentiones super hoc*": Statuta Communis Parmae (1855), II, p. 39. In fact, beyond factions, sortition was also intended to reduce corruption and violence: see Wolfson (1899), p. 12.

Intentional selection of individuals for public office was simply placed alongside sortition by lot. The sheer sum of individual wills, to be turned into collective will, would not yield a sensible outcome. Sortition would add to the conscious choice of representatives something that choice could not offer. Sortition is neutral, as it rules out persuasion. It also implies no responsibility, as it rules out mandate. Both elements could prove to be useful.

It must be stressed that in the cases mentioned before sortition by lot – and thus the unpredictability of outcomes - has in actual fact contributed to reduce factionalism. The frequent turnover of members of bodies like the Athenian Boule or the Florentine Signoria made it very difficult for individual members to establish a reputation. Quite clearly this made the discussions and negotiations necessary to reach collective agreements rather lengthy and difficult. In economic parlance this implied high transaction costs. The practice of log-rolling, i.e. the exchange of favours, was as a result of such difficulties unlikely to develop, both within and among factions. Thus, by weakening the likely bonds to be established within each political group, sortition has done a good job in contrasting factionalism, usually considered detrimental to orderly public life. However, in contemporary parliaments the longer political terms and the possibility of running for the same office more than once has, with the aid of party discipline, reduced transaction costs but increased the practice of log-rolling.

The mathematical model we are going to present in the next sections displays members of parliament distinguished according to whether they are subjected or not to party discipline. In the model party discipline is taken to represent all the features of a system prone to factionalism, like long political terms or running more than once for the same office. In the model, party discipline implies that all members of the group follow the party leader in all their decisions. In case party discipline does not apply, everybody will decide independently of each other and factionalism is put into check. Freeing members of parliament from party discipline, therefore, helps lot yield all its beneficial effects.

**3. Modelling a parliament**

We assume that voting is the way individuals take collective decisions. In the case of a parliament (or any other deliberative body) those decisions are acts of parliament. The problem we are trying to address here is whether the whole job of making decisions (the job of members of parliament) can yield better results by changing the composition of parliament. To this end, prospective decisions (or proposed acts of parliament) are ordered according to the social gain they are capable of generating, from the least beneficial ones (or the most harmful) to the most beneficial ones. Decisions are submitted to parliament. In actual fact, only those decisions that command a majority of votes are taken. Since the bunch of decisions taken at any given run of parliament may differ in the aggregate social gain it manages to generate, it is interesting to ask whether different compositions of parliament may have a differential impact on that aggregate social gain.

There is no discussion in the following on the way such an ordering is set up. It is assumed right at the beginning that it is a socially accepted ordering. Therefore it is accepted by members of parliament, as well. However, members of parliament may have a different idea as to the threshold that makes a given proposal acceptable or not. In the following it is assumed that members of parliament are distributed equally across the socially accepted ordering of decisions, so that we have members who would be prepared to accept anything, however low may be the social contribution the decision is supposed to make, and members of parliament who expect decisions to produce a significant contribution to social welfare before they can be prepared to vote in their favour.

The way proposals are submitted is not going to affect the way proposals are voted. This means that submission could be modelled in many different ways. It is assumed here that it is legislators who take the responsibility for submitting proposals. They are expected to submit the same number of proposals. It goes without saying that individual proposals must meet the threshold associated to each legislator. In this particular case it is assumed that the threshold is just met. Legislators submit proposals that are expected to generate a social gain no larger or smaller than their personal welfare threshold.

A well functioning parliament is one which, during any single term, can be expected to yield the best possible results, as measured by the aggregate social welfare gain. Ideally, parliaments should be prepared to pass only those acts that produce positive contributions to welfare. In fact, members of parliament may be prepared to accept even acts that produce negative contributions to welfare. If a majority for those acts can be put together, those acts will be passed, thus producing a negative contribution to social welfare. The following analysis is designed to look for those conditions required to make sure that parliaments reduce the risk of making negative contributions to welfare.

In the following we first present a one dimensional mathematical model of a parliament with $N$ members (legislators) $A_i$ (i=1,...,N) and two parties (or coalitions), the majority party ($P_1$) and the opposition (minority) party ($P_2$). From now on we will consider the terms "party" and "coalition" as synonymous, since our model applies indistinctly to both of them. For the time being we set the percentage of legislators of the two parties as: $P_1$ = 60% and $P_2$ = 40%.

During a given Legislature L, each legislator can perform only two actions: submitting a bill for approval and voting in favour or against any bill. Legislators are represented as points of a 1D space, i.e. of an horizontal axis indicated with the capital letter Y. Each point of this Y-axis, associated to a real number in the interval [-1,+1], shows the (average) social gain attached to the bill submitted by the legislator $A_k$. As mentioned above, this represents the welfare threshold below which the legislator is not prepared to go when it comes to vote. In a previous study (A. Pluchino et al., 2011) we considered a 2D model of parliament, where two different thresholds were associated to each legislator, one related to a personal interest (X-axis) and another related to the social welfare (Y-axis). In the 1D version employed here, to make the model simpler and more suitable for an analytical approach, we explicitly consider only the second threshold (maintaining the same name for the axis), but somehow account for the influence of the personal interest (as explained later).

Within the 1D space, the distributions of legislators belonging to the two parties can be represented, in the limit for N>>1, as two probability density functions (PDF) defined over the Y-axis (see Figure 1). Both the distributions $P_1(Y)$ and $P_2(Y)$ are assumed to be Gaussians, with means $<Y_1>$ and $<Y_2>$, and with the same standard deviation $\sigma$. Both the curves are normalized to have unitary area, but the size of the $P_2$ curve in this and in the next figures has been reduced only in order to distinguish it at a first sight.

For a given legislative term L, the centroids $<Y_1>$ and $<Y_2>$ of the distributions are fixed and randomly chosen in the interval $[-1 + \sigma, 1 - \sigma]$. During each parliamentary term (legislature), we assume that each legislator puts forward the same number of proposals. Therefore, the overall percentage of proposals coming from the members of a particular party is equal to the percentage of legislators of that party. Since the Y coordinate of each legislator represents the social gain of her proposal, the distributions $P_1(Y)$ and $P_2(Y)$ can be also regarded as the distributions of their proposals.

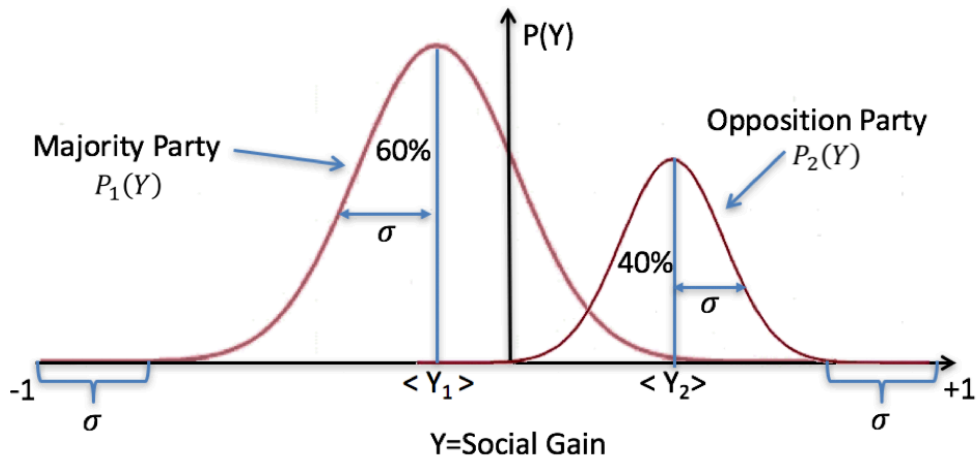

Figure 1. Gaussian distributions of legislators belonging to the two parties $P_1(Y)$ and $P_2(Y)$. The two curves also represent the distributions of proposals coming from the two parties.

When it comes to voting a given proposal, say p* with abscissa Y*, party discipline applies for each legislator $A_k$. Suppose $A_k$ belongs to the majority party $P_1$ (but of course the same considerations hold also for $P_2$), there are two possibilities:

- p* is an internal proposal (i.e. it comes from a member of $P_1$): it is accepted by $A_k$ regardless of its social gain Y*;

- p* is an external proposal (i.e. it comes from a member of the other party $P_2$): it is accepted only if Y* is greater than the party mean $< Y_1 >$, which represents the *minimum* social gain which a proposal coming from legislators of $P_2$ should yield to be accepted by legislators of $P_1$; moreover, if this condition is fulfilled, we assume that, due to internal motivations (personal interests) of $P_1$, only 50% of its legislators will accept the proposal.

Proposals are accepted by parliament provided they receive half plus one ($N/2 + 1$) of the votes. Due to party discipline for internal proposals, this requirement will be always fulfilled for the majority party $P_1$, which alone owns 60 per cent of the legislators. We should not forget, however, the proposals of the opposition party. The final number of accepted proposals at the end of a term will depend also on the contribution of the opposition Party $P_2$, whose proposals are voted inasmuch as the majority party likes them. Therefore, the number of opposition party proposals, which finally gets to be approved, should also depend on the relative position of the two parties along the Y axis.

Actually, as already mentioned, due to the influence of personal interests, only *one half* of members of $P_1$ will vote for the opposition proposals lying on the right of the $P_1$ mean. This means that, for these proposals (indicated by the dark parts of the $P_2$ Gaussians in Figure 2), 30% of legislators belonging to $P_1$ will sum their vote to the 40% of legislators belonging to $P_2$, thus always exceeding the $N/2+1$ threshold necessary for the approval of the proposals.

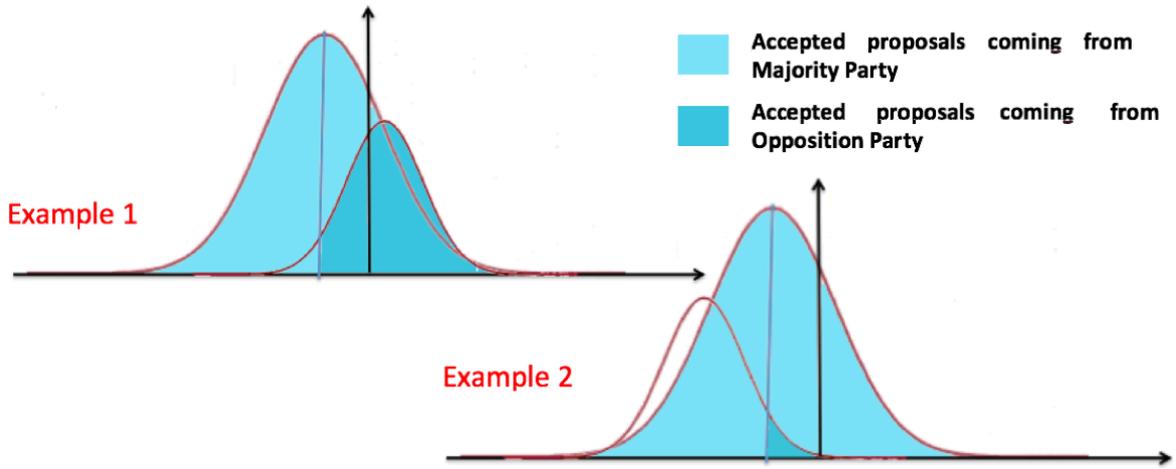

Figure 2. A couple of example where the fraction of accepted proposals coming from $P_2$ (in dark blue) are shown as function of the relative positions of the two parties.

In this paper we will always consider a sequence of $N_L$ parliamentary terms, each one with a different random position of the centroids of both parties. It is reasonable to assume that, averaging over many terms ($N_L \gg 1$), the asymptotic distributions of both parties will be centred at Y=0 (see Figure 3). In the following we will always make this assumption for all the analytic derivations. In particular, under this hypothesis, the opposition party proposals, which finally get to be approved with the contribution of the majority, will be all the positive ones. We will analyse in detail this situation in the next section.

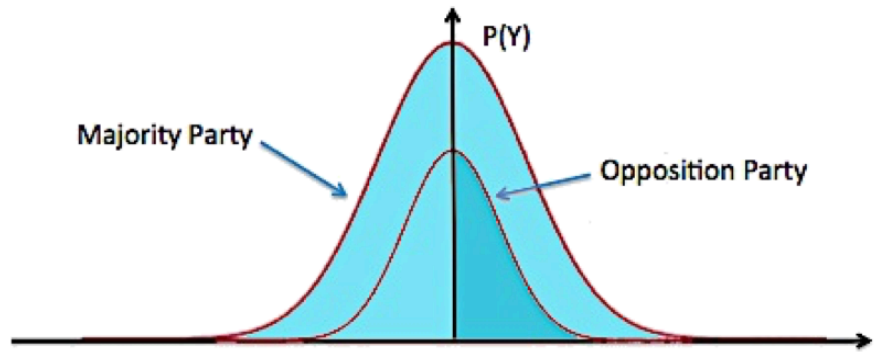

Figure 3. Asymptotic distributions of both parties after averaging over many terms ($N_L \gg 1$). The contribution of the opposition party $P_2$ to the expected percentage of accepted proposals is now represented by the *positive* side (in dark blue) of the corresponding asymptotic distribution.

Let us define, now, the global efficiency *Eff* (L) of a term. It will be given by the net social gain yielded by the accepted proposals, i.e. by the product of the percentage $N_{\%ACC}$ (L) of the accepted proposals (calculated with respect to the total number of proposals put forward during the term) times their average social gain $Y_{AV}$ (L):

$$Eff(L) = N_{\%ACC}(L) \cdot Y_{AV}(L) \qquad (1)$$

Averaging over $N_L$ terms, one can obtain the expected global efficiency:

$$Eff_{exp} = \frac{1}{N_L}\sum_{j=1}^{N_L} Eff(L) \qquad (2)$$

Defining $N_{\%ACC}$ as the expected percentage of accepted proposals and $Y_{AV}$ as their expected social gain, both averaged over the $N_L$ terms:

$$N_{\%ACC} = \frac{1}{N_L}\sum_{j=1}^{N_L} N_{\%ACC}(L) \qquad Y_{AV} = \frac{1}{N_L}\sum_{j=1}^{N_L} Y_{AV}(L)$$

one can also expect that, in the limit $N_L \gg 1$:

$$Eff_{exp} \cong N_{\%ACC} \cdot Y_{AV} \qquad (3)$$

The expected global efficiency is the fundamental measure of the efficiency of a deliberative body like a parliament. In the following we shall develop this notion much further with the purpose of investigating the effects on such a measure of a variable composition of parliament. In particular, we shall address the question of what would happen if a variable percentage of members of parliament had no longer an obligation of following the party line, simply because they do not belong to any party. For this reason, these legislators will be called "independent". At variance with legislators elected as members of a party,[12] independent legislators might be assumed to be drawn randomly from the population of constituents. The whole purpose of this paper is precisely to show how the expected global efficiency changes as parliament allows a given number $N_{ind}$ of randomly selected independent members in, free of any party discipline.

## 4. Expected global efficiency for the two polar cases: $N_{ind} = 0$ and $N_{ind} = N$

In this section we will show how to derive analytically the proposed measure of parliamentary efficiency in the limit of many legislators ($N \gg 1$) and many parliamentary terms ($N_L \gg 1$). These limits will allow us to consider Gaussian distributions of legislators and to substitute integrals to the summations for calculating the averages of the various quantities. For the time being, the derivation will concern the two polar cases: a parliament with just *two parties*, i.e. with $N_{ind} = 0$ independent legislators, and a parliament with just *randomly selected members*, i.e. with $N_{ind} = N$.

*A Parliament with only two parties (60%-40%)*
To determine the global efficiency in the case of a parliament with a majority party $P_1$ with 60 per cent of members and an opposition party $P_2$ with 40 per cent, we proceed by evaluating separately the two factors in Equation 3. In accordance with the assumption $N_L \gg 1$, for the Gaussian distributions of the two parties we assume that $<Y_1>=<Y_2>=0$. We also assume the same standard deviation $\sigma=0.15$. Because of the previously illustrated voting rules and recalling that the percentage of proposals coming from the members of a party is equal to the percentage of legislators of that party, the expected percentage of accepted proposals $N_{\%ACC}$, averaged over many parliamentary terms, is given by the sum of two elements (consider again Figure 3 as reference): the first one, which is the contribution of party $P_1$, is represented by the *whole* area below the corresponding asymptotic distribution $P_1(Y)$, while the second one, the contribution of party P₂, is represented only by the (dark) area below the *positive* part of the corresponding asymptotic distribution P₂(Y), since – as already shown in the previous section – the positive proposals coming from the opposition party are the only ones voted also from the majority party.

---

[12] As a matter of fact no electoral system is explicitly assumed in the paper. The particular distribution of party members however is an indication that those members of parliament are not randomly selected.

As shown in Appendix A, in this case the expected global efficiency over many parliamentary terms will be:

$$Eff_{exp} = N_{\%ACC} \cdot Y_{AV} = 80(\%) \cdot 0.06 = 4.8\ \% \tag{4}$$

It is not surprising that we get such a very small value. The only positive contribution to social welfare comes from the opposition party that, thanks to the support of the majority party, will see its positive proposals approved with a large majority. It may sound paradoxical that the majority party is capable of giving rise to a positive social gain only when other parties' proposals are at stake. We will see later that this value can be increased if different circumstances apply. In particular, we shall look at the possibility of filling parliament with independent legislators, free from any party linkage.

This could be realized in practice by selecting them at random from all the citizens with the necessary requirements (in principle, the same that allow them to express their preferences for the parties during the elections). Because of this particular selection procedure, we will assume that the independent legislators are not subject to any kind of party discipline: each of them votes independently from the other independent legislators and from the parties. In particular, given a proposal p* with abscissa Y*, any independent member $A_k$ with abscissa $Y_k$ will accept it only if Y*> $Y_k$, since $Y_k$ represents the *minimum* social gain that the proposal should yield for it to be accepted by $A_k$. Again, as already seen for the parties' vote, for a given proposal p*, we assume that, because of internal motivations (*personal interests*), only 50% of the independent legislators fulfilling the condition $Y_k$ < Y* will accept the proposal.

*A Parliament with only independent legislators*

Let us now consider the extreme case of an entirely independent parliament, that is a parliament with only independent legislators ($N_{ind}$ = N). We assume that, in the limit N>>1, their probability distribution P(Y) along the Y-axis is a uniform one (from -1 to +1) with unitary area (see Figure 4). Due to this latter requirement, we will have P(Y)=1/2.

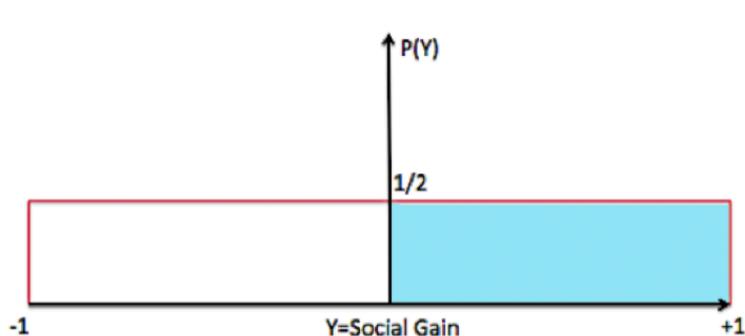
Figure 4. Uniform distribution of independent legislators

We know that the main feature of independent legislators is that they are not subjected to any party discipline. This, in principle, brings into the parliament a positive element, since each proposal needs to be largely discussed to reach the consensus of the majority (N/2 + 1) of legislators. However, in such circumstances (since an average over many parliamentary terms is considered), none of the proposals will be accepted. In fact, it is quite clear that only proposals with a positive abscissa Y* could have a chance to be accepted, since the required majority could be reached only for those proposals (see dark area in Figure 4). However, because of their personal interests, we know that only one half of the independent legislators with $Y_k < Y^*$ will accept these proposals. Therefore, to reach $N/2 + 1$ votes, a given proposal should have an abscissa Y* not only greater than zero but also greater than 1.

In conclusion, for a hypothetical parliament with only independent legislators, we will always find a null result for the expected values of both the percentage of accepted proposals and the average social gain. This means that the expected global efficiency over many terms will be:

$$Eff_{exp} = N_{\%ACC} \cdot Y_{AV} = 0 \qquad (5)$$

This unexpected result makes party discipline not so difficult to accept, especially when it makes good proposals (i.e. with a positive value of social gain) easier to accept. But certainly not when it is designed to impose the dictatorship of the majority. As argued above, in a parliament without parties, none of the proposals has enough votes to get approved, since no proposal is good enough for half of the legislators. This result mirrors that arising from a parliament with only parties, where hardly any positive contribution to social welfare is likely on average to emerge.

However, it has certainly not gone unnoticed that the very small efficiency associated to both illustrated cases - a parliament with only parties and one with only independent legislators - depends on opposite reasons. In the case of a parliament with parties many proposals get accepted, but they yield a very small average social gain. In the case of a parliament with only independent legislators, parliament accepts only extremely good proposals, but their number is close to zero. Therefore, we can expect that contaminating a parliament with two parties with an increasing number of independent legislators would reduces the number of accepted proposals while increasing their average social gain (at least until the percentage of accepted proposals is greater than zero). As a consequence, the value of the global efficiency as a function of the percentage of independents sneaking into parliament, calculated as product point by point of the previous two quantities, should show an initial increase from its quite low extreme value typical of the two-party case, then it should reach a central peak in correspondence of a given percentage of uniformly distributed independents, and finally it should slip towards zero, when the percentage of independents approaches a hundred per cent. In the following section we will provide a demonstration that this is roughly the case.

**5. A parliament with an increasing number of independent members**

Let us start with the analytic derivation of the percentage of accepted proposals as function of the number of independents, in a parliament with *N>>1* members and in the limit of many legislative terms. Just as an example, we will consider a parliament with *N* = 500 members, but of course our results are valid for any (great) value of *N*. It is worth noticing that, with $N_{ind}$ independent legislators, the real number of members belonging to the two parties, $P_1$ and $P_2$, has to be calculated by taking, respectively, 60 per cent and 40 per cent of the difference $N - N_{ind}$. This means that, above a given threshold of $N_{ind}$, $P_1$ will no longer be the *absolute* majority party but only the *relative* majority one. We will see that it is precisely this feature, together with the role of the independent legislators that makes enhancing the efficiency of the parliament possible.

As we know, in the limit of many legislative terms, both parties are centred at Y=0, thus positive proposals (i.e. proposals with Y* > 0), which represent the 50% of the total number, are accepted by 100 per cent of the members of the party proposing them (through party discipline) and by 50 per cent of the members of the other party (due to the influence of personal interests). Furthermore, in principle, these proposals should be accepted also by one half of the independent legislators with abscissa Y < 0. But, from the independent legislators point of view, the average social gain of the positive proposals is not Y=0 but Y=0.5 (the middle point of the positive part of the Y axis). One can easily notice that some proposals, lying on the positive axis, will be voted by independent legislators having an abscissa higher than zero. Therefore, one half of the independent ones with -1 < Y < 0.5 will vote, on average, for positive proposals: this means that, since they are uniformly distributed along the Y axis, a percentage of 37,5% of independent legislators will accept these proposals (as shown in Figure 5).

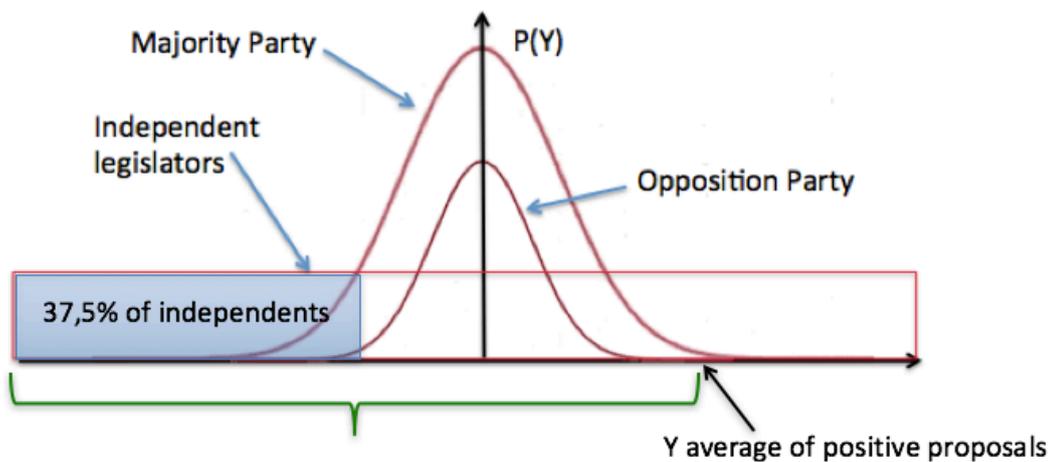

Figure 5. Percentage of independent legislators voting, on average and in the limit of many legislative terms, for the positive proposals coming from the parties

Pooling together the three contributions of the majority party, the opposition party and independent legislators, we can easily work out the number of votes received, on average, by the positive proposals coming from the legislators, as a function of the percentage of independent legislators sitting in parliament. The results are shown in Figure 6.

Looking at the details of the plot, where the example of a parliament with N=500 members is considered, one can notice that:
- positive proposals coming from independent legislators are never accepted, since they never reach the threshold of $N/2 + 1 = 251$ votes (indicated by the horizontal line);
- positive proposals coming from the relative minority party $P_2$ are accepted until the percentage of independent legislators in the parliament stays below 61 per cent;
- positive proposals coming from the relative majority party $P_1$ are accepted until the percentage of independent legislators in the parliament stays below 70 per cent.
-

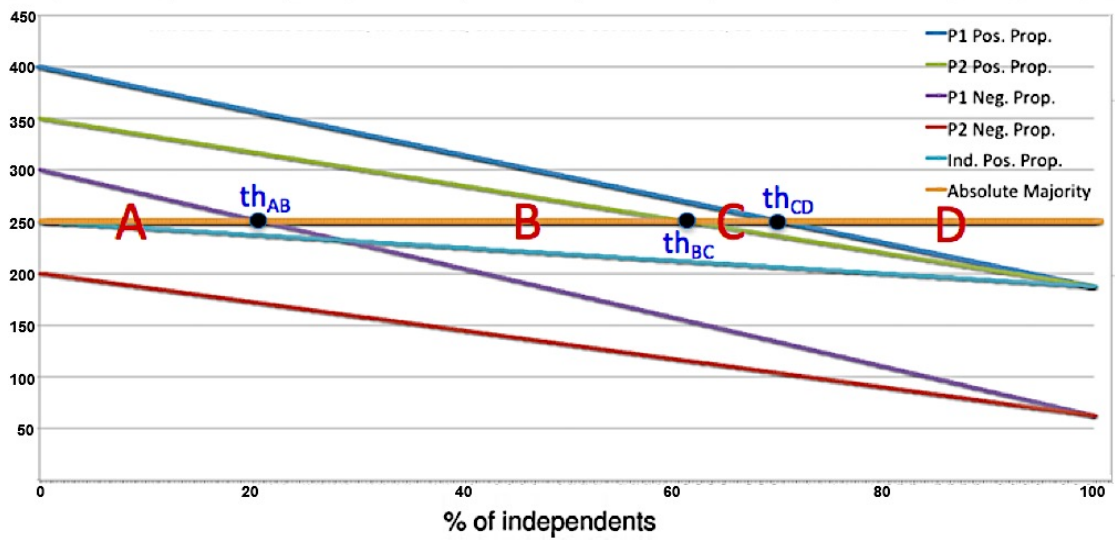

Figure 6. Number of votes received, on average, by both the positive and the negative proposals coming from the three component of the parliament, as a function of the percentage of independent legislators. The absolute majority level is also reported as a horizontal line. Along this line, three thresholds are visible, under which the corresponding proposals do not reach enough votes to get passed.

In the same Figure, negative proposals have been also considered (i.e. proposals with $Y^* < 0$). It is straightforward to notice (and it is confirmed by the plot) that only those coming from the relative majority party $P_1$ can be accepted, until this party maintains the absolute majority, since its members accept both good and bad proposals because of party discipline. Negative proposals coming from $P_2$ and independent legislators are never accepted. When $P_1$ loses the absolute majority, its negative proposals (on average) can still be accepted due to the contribution of one half of independent legislators with $-1 < Y < -0.5$ (who represent the 12,5% of the total, see Figure 7), being $Y = -0.5$ the average social gain of negative proposals coming from $P_1$. This happens until the percentage of independent legislators in the parliament stays below 21 per cent. It is interesting to notice that neither positive nor negative proposals coming from independent legislators can be accepted, no matter how many of them sit in parliament.

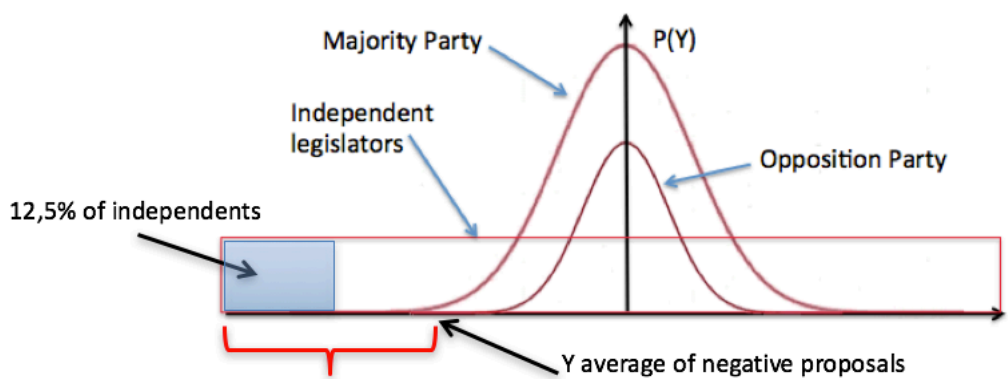

Figure 7: Percentage of independent legislators voting, on average and in the limit of many legislative terms, for the negative proposals coming from the parties.

Summarizing, in Figure 6 we saw that three progressive thresholds $th_{AB}$, $th_{BC}$ and $th_{CD}$ do exist, identifying four different intervals in the percentage of independent legislators, namely A, B, C and D, each one with a given (decreasing) percentage of accepted proposals. Quite clearly, the values of these thresholds strictly depend on the size *p* of the relative majority party $P_1$. In the case considered, i.e. for *p* = 60 per cent, we already found (empirically) $th_{AB}$=21% $th_{BC}$=61% and $th_{CD}$=70%. The expressions of these thresholds for any value of *p* and any value of N have been analytically determined in Appendix B.

Having a look at the behaviour of $th_{AB}$, $th_{BC}$, $th_{CD}$ as function of *p* for *N*=500, reported in Figure 8, the four regions A, B, C and D are well defined in a plausible range of values of *p*, going from 51% to 80%. On the other hand, for *p* > 80% the region B would disappear, but the values of *p* would be unrealistic: in fact, also with randomly selected legislators, the relative size of the two parties in the real world should continue to be decided by elections results (in this respect we imagine a *mixed* electoral system), and it is very unlikely that any party or coalition obtains more than 60-65% of preferences.

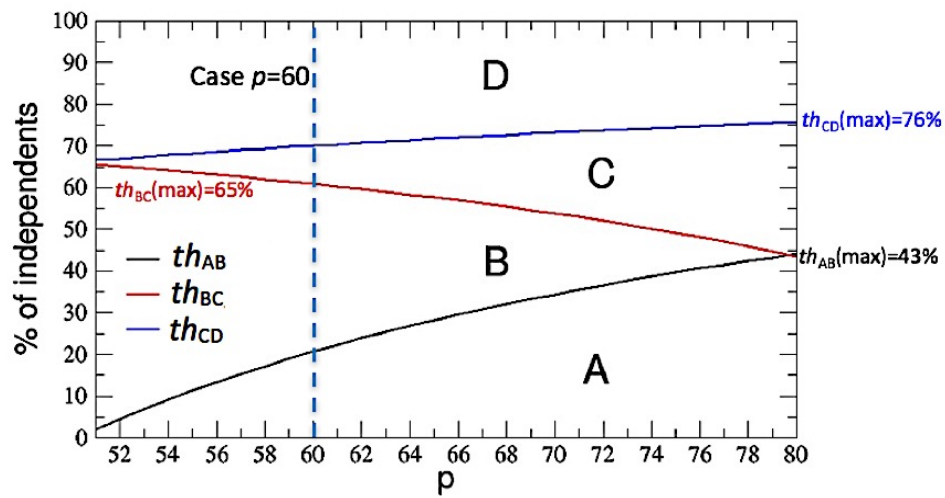

Figure 8. Behaviour of the three thresholds $th_{AB}$, $th_{BC}$, $th_{CD}$ as function of *p* for *N*=500. The particular case of *p*=60% is reported as a dashed vertical line.

At this point, within each one of these four regions, we can analytically work out, in the limit of many legislative terms, the percentage of accepted proposals $N_{\%ACC}$ and the average value of the social gain $Y_{AV}$ produced by these proposals, both as a function of the number of independent legislators $N_{ind}$. Details of this derivation are in Appendix C.

## 6. Searching for the maximum efficiency

We are now ready to calculate the expected average global efficiency $Eff_{exp}$ of our parliament as a function of p, N and $N_{ind}$, in the four regions A, B, C and D, by multiplying inside each of them the value of $N_{\%ACC}$ (Appendix C, Equations 1-4) for the corresponding value of $Y_{AV}$ (Appendix C, Equations 5-8). Details of the calculations are in Appendix D. Let us present, here, the plots of these efficiency functions into each region.

In Figure 9 we plot the efficiency $Eff_A$ (Appendix D, Equation 2) as function of $N_{ind}/N$ for three increasing values of *p* from 51% to 70%. We observe that, within its maximum possible range (i.e. between 0 and $th_{AB}$(max)=43%, i.e. for 0 < $N_{ind}/N$ < 0.43), it is a monotonically increasing function (as also confirmed by the derivative plot in the inset):

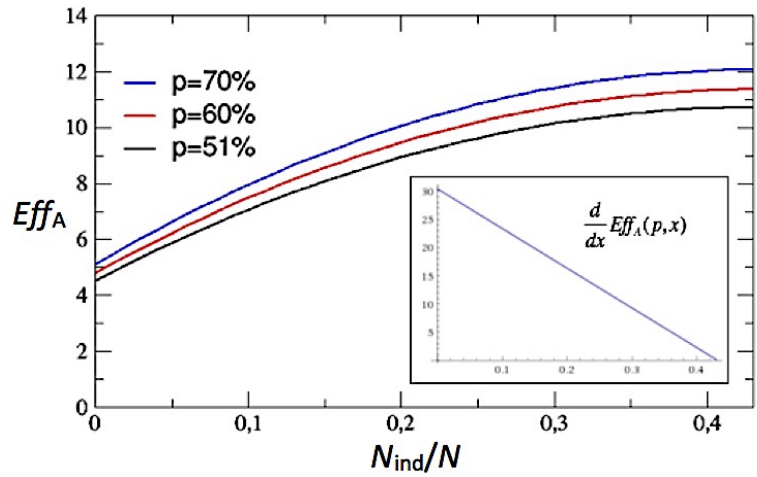

Figure 9: Behaviour of the efficiency $Eff_A$ as function of $N_{ind}/N$ for three increasing values of $p$. The derivative of $Eff_A$ is plotted in the inset and it is always positive in the same interval for any $p$.

In Figure 10 we plot the efficiency $Eff_B$ (Appendix D, Equation 3) as function of $N_{ind}/N$, which does not depend on $p$. We observe that, in its maximum possible range (i.e. between 0 and $th_{BC}(max)=65\%$), it has always a maximum.

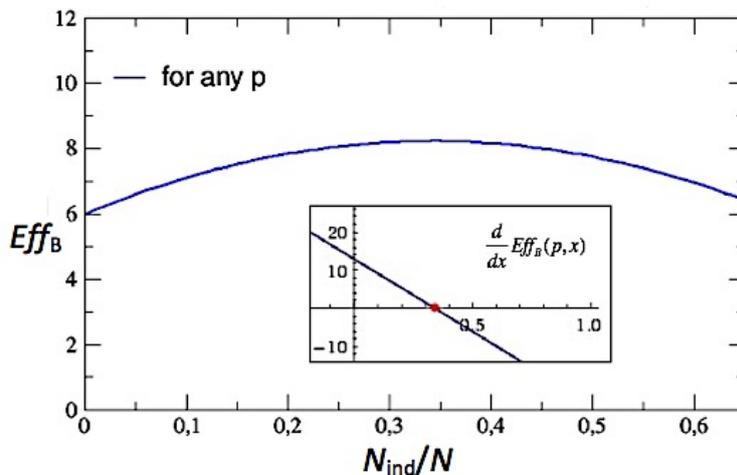

Figure 10. Behaviour of the efficiency $Eff_B$ as function of $N_{ind}/N$ for any value of $p$. The derivative of $Eff_B$ is plotted in the inset and the position of the maximum is visible.

The value of the maximum ($N_{ind} = 34\%$, see Appendix D) ensures that it remains in region B until the size $p$ of the majority party stays below 70% (above this size, as visible in Figure 8, region B becomes too smaller).

Plotting in Figure 11 the efficiency $Eff_C$ (Appendix D, Equation 4) as function of $N_{ind}/N$ for the same three values of $p$ considered in Figure 10, we notice that, in its maximum possible range (i.e. between $th_{AB}(max)=43\%$ and $th_{CD}(max)=76\%$), it is a monotonically decreasing function, as also confirmed by its derivative plotted in the inset (see Appendix D).

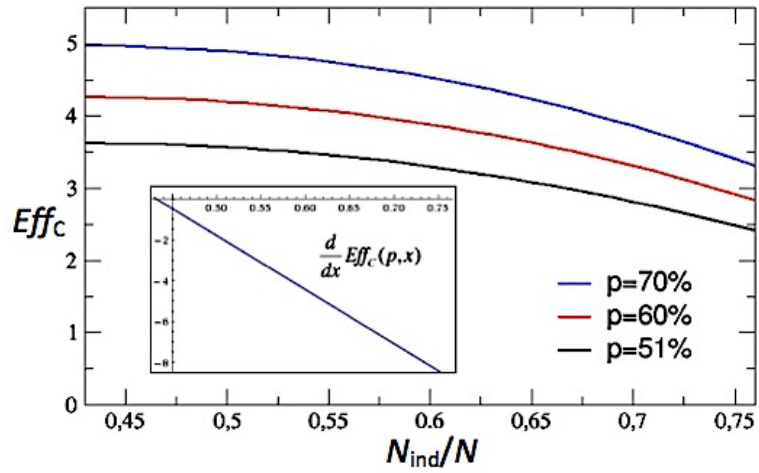

Figure 11. Behaviour of the efficiency $Eff_C$ as function of $N_{ind}/N$ for three increasing values of $p$. The derivative of $Eff_C$ is also plotted in the inset and it is always negative in the same interval for any $p$.

Summing up so far, the behaviour of the average global efficiency in the three regions A, B and C, where it assumes non null values, seems consistent with the hypothesis that this efficiency has a minimum at the two extrema ($N_{ind}=0$ and $N_{ind}=N$), starts to monotonically increase for small values of $N_{ind}$ (region A), reaches a maximum for $N_{ind}/N=0.34$ (region B), then monotonically decreases towards zero (region C). From there on, in region D, the efficiency $Eff_D$ remains equal to zero (see Appendix D, Equation 5).

Let us now look into this scenario in greater detail, and plot in Figure 12 the average global efficiency in the range $0 < N_{ind} < th_{CD}(max)$ for four increasing values of $p$. The positions of the three thresholds $th_{AB}$, $th_{BC}$ and $th_{CD}$ that separate the four regions are indicated by vertical dashed lines. Of course, the sudden change in efficiency observed in all the plots when the value of $N_{ind}/N$ crosses each one of the three thresholds, is an effect of the assumed limit of many (infinite) legislative terms. Averaging over a relatively small number of them, the fluctuations mainly due to the random positions of the two parties along the Y axis would make these transitions much smoother.

By looking closely at the four panels, it clearly appears that the position of the maximum value for the global efficiency, say $Eff_{max}$, is not always situated in region B but strictly depends on the value of $p$. As is visible in panels (a) and (b), the initial insertion of independent legislators in a parliament with only parties induces a sudden increase in the global efficiency, similar for any value of p, which reaches its maximum value $Eff_{max}(A)$ at the threshold $th_{AB}$. However, since the position of this threshold does depend on $p$, for intermediate values of $N_{ind}$ it may occur that, as shown in panels (c) and (d), the value $Eff_{max}(A)$ exceeds the maximum value of the efficiency in region B, $Eff_{max}(B)$.

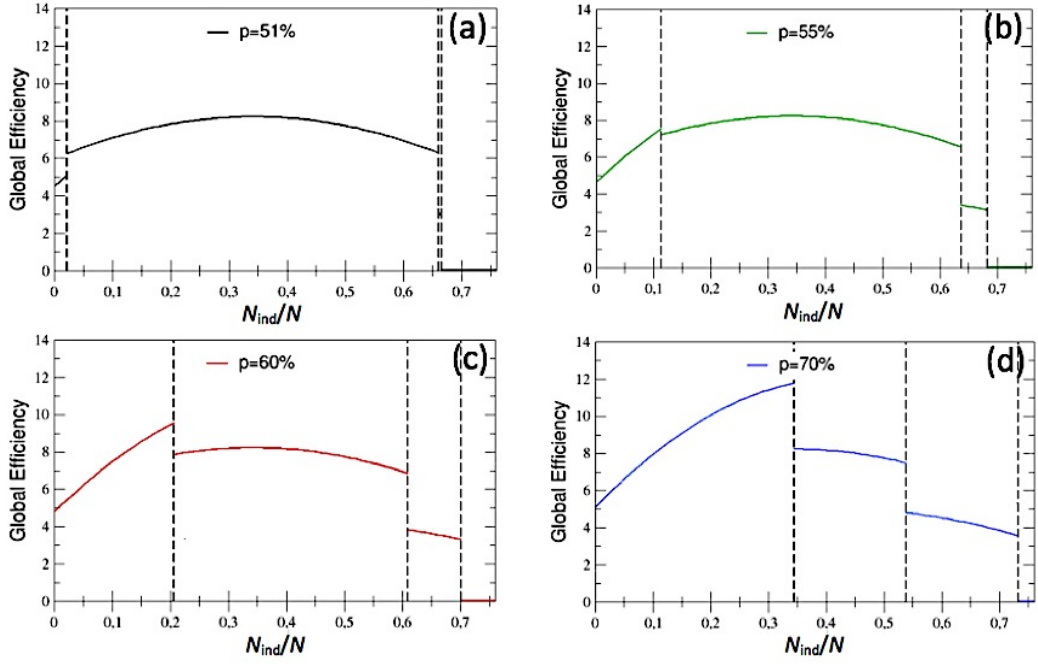

Figure 12. The average global efficiency in the range $0 < N_{ind} < th_{CD}(max)$ is plotted for four increasing values of $p$, i.e.: 51% (a), 55% (b), 60% (c) and 70% (d).

Following these insights, in Figure 13 we plot, as dashed lines, both the constant position of the absolute maximum for $Eff_{max}(B)$ (34% of independent legislators) and the variable position of the threshold $th_{AB}$, also expressed as percentage of the independent legislators. Then, in bold, we highlight the position of the global maximum efficiency $Eff_{max}(p)$. We found that, until $p < 56.5$ (%), it results $Eff_{max}(B) = 8.22 > Eff_{max}(A)$, therefore $Eff_{max}(p) = Eff_{max}(B)$. However, for $p > 56.5$ (%), $Eff_{max}(A)$ starts to exceed 8.22, thus becoming the new global maximum efficiency $Eff_{max}(p)$. This implies that at $p = 56.5$ (%) the percentage of independent legislators needed to get the maximum efficiency suddenly rushes down from 34% to 14.3%, then it slowly goes back towards 34%, reached again around $p = 70$ (%), where region B tends to disappear and the maximum efficiency reaches its highest value $Eff_{max}(A) = 11.8$.

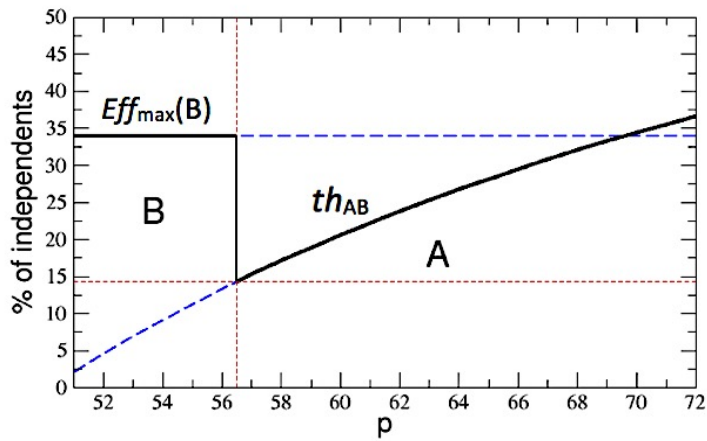

Figure 13. The position of the absolute maximum for $Eff_{max}(B)$ and the variable position of the threshold $th_{AB} = Eff_{max}(A)$ are plotted as dashed lines. In bold, partially superimposed to the previous lines, we indicate the position of the global maximum efficiency $Eff_{max}(p)$.

In conclusion, the analysis of the parliament in the limit of many legislative terms confirms that an intermediate percentage $N_{\%ind}$ (between 14% and 34%) of legislators independent from the two parties and not subject to any party discipline, can always improve the efficiency of the system, regardless of the size of the relative majority party. This general analytical result is in full agreement with our previous findings, obtained through a computational study of the 2D model of parliament, where both a variable relative size for the two parties and a variable percentage of independent legislators were considered [A. Pluchino et al., 2011] and suggests a possible policy for effectively increasing parliamentary efficiency.

**7. Repairing the fallacy of representative democracy**

Representative democracy, when based on the party system, fails to deliver its most important outcome, i.e. an efficient bundle of acts of legislation. This is especially the case when the majority representation system is adopted. In this case, there is always a party in a position to command an absolute majority in parliament. This system is usually advocated as a system capable of delivering decisions, thus denying any serious fallacy and arguing that the ability of taking decisions makes up for the failings of representation and party discipline. In fact, one can improve upon that outcome. This paper was designed to show that this is possible when independent legislators, randomly selected from the relevant constituency, are allowed to sit in parliament without any obligation to join a party and accept its discipline.

The main argument was based upon the costs a representative democracy typically entails. Those costs emerge from the failure of the relationship between constituents and political representatives. It typically happens that political representatives cease to be faithful agents of their constituents and turn into agents of their parties. This is what we have called factionalism. As extensively illustrated in the paper, this shift in the agency relationship deprives the representative system of its ability to deliver decisions that command a large consensus in the population of constituents. For one can expect that the system delivers unpopular decisions just as easily as it can deliver popular ones. Its cost, therefore, is the forgone benefit of not delivering just popular decisions.

It has been argued in the paper that there is a way to repair this fallacy and minimize the cost of representative democracy. The logic underlying this endeavour is preserving the ability of representative democracy to deliver decisions while increasing as much as possible their net contribution to welfare. The way to do so is shifting the balance of decision making towards the positive side. Letting a number of independent legislators, drawn at random among common citizens, sitting in parliament with the same prerogatives of the other legislators, has an outstanding advantage: it allows in parliament legislators who will vote only for acts that meet their personal threshold. This is a major improvement in a system where legislators, due to party discipline, vote also for acts they do not like. By gradually allowing more and more independent legislators in parliament, the detrimental effects of party discipline are curtailed and the beneficial effects of voting according to one's preferences are enlarged. There comes a point where the ability of a party to make its members vote for whatever comes from within loses its strength, thus magnifying the role of those legislators who are not subjected to any imposition. However, one should not think that the larger the number of independent legislators, the better. Party discipline has its merits and should not be discarded altogether. As extensively shown in the previous sections, if only independent legislators sat in parliament hardly any decision would be taken, whether good or bad. The model of parliament illustrated in this paper has precisely shown that a virtuous combination of party discipline and freedom of choice is possible.

Implementing a mixed parliament is not impossible. For example, the number of independent legislators to be introduced in the parliament could be linked to the level of abstention in the election (in the 2014 European elections the abstention area was by far the first party, with a 57% of non voters, but a typical percentage in almost all modern democracies is around 30%). Our proposal may provide an option to those constituents who are typically oriented towards abstention: each of them, going to the polls during the election day, could choose whether to vote for a party candidate or enrol in a list for a sortition. Then, after the elections, a percentage of seats proportional to the area of abstention would be reserved to randomly selected citizens picked out from the list of candidates. Of course the remaining seats would be assigned to candidates of the parties in a proportion established through voting.

It is likely that such a procedure will give rise to a parliament without any absolute majority, and this could be considered dangerous in parliamentary systems, since it could make difficult to have a stable Government (which is usually expressed by the absolute majority). But, as we show in the paper, the absence of an absolute majority, because of independent legislators, is precisely what improves the efficiency of the parliament. Finally, to protect the independent legislators from being captured by the existing parties, we could also think of a system of rotation, so that new independent legislators would be selected at random (from the original sortition list decided by the abstention level) for each single parliamentary session, devoted to a specific issue.

## 8. Concluding remarks

The purpose of this paper was to show that sortition can help reduce the detrimental effects of factionalism. In the paper those detrimental effects, which emerge from party discipline, are foregone benefits. When two or more parties sit in a parliament, party discipline on average generates grossly inefficient results, as good decisions tend to be cancelled by bad decisions. If, on the contrary, a number of members of parliament sorted by lot are introduced, party discipline does not spread across parliament and that null average can be turned into a positive outcome. The paper shows that there are thresholds beyond which improvements are no longer possible. That is the ideal state of affairs. Just like in old Italian city-states when sortition could be effectively combined with more traditional ways of selecting public officers, nowadays sortition could be combined with party systems to tame the undesirable implications of factionalism.

**APPENDIX A**

Here follows the formal expression:

$$N_{\%ACC} = \left[\int_{-1}^{1} P_1(Y)dY\right] \cdot 60\% + \left[\int_{0}^{1} P_2(Y)dY\right] \cdot 40\%$$

where the two contributions are weigthted according to the percentage of proposals put forward by each party, in this case equal to its relative size. Since both the distributions are normalized, i.e. their total area is equal to 1, the expected value for $N_{\%ACC}$ over the entire set of $N_L$ terms will be:

$$N_{\%ACC} = 1 \cdot 60 + 0.5 \cdot 40 = 80\%$$

In the same fashion, the expected average social gain $Y_{AV}$ of these proposals over the same set of parliamentary terms is given, again, by the sum of two elements, stemming from the contributions of $P_1$ and $P_2$ (in terms of fraction of proposals submitted multiplied by their welfare contribution)

$$Y_{AV} = \left[\int_{-1}^{1} P_1(Y) Y\, dY\right] + \left[\int_{0}^{1} P_2(Y) Y\, dY\right]$$

The first integral gives a null result since, because of party discipline, the majority party accepts all its internal proposals, regardless of their contribution (positive or negative) to the social gain; on the other hand, the second integral is calculated only over the positive Y-axis, therefore it gives a small positive result:

$$Y_{AV} \approx 0 + 0.06 = 0.06$$

Therefore, the expected global efficiency over many parliamentary terms will be:

$$Eff_{exp} = N_{\%ACC} \cdot Y_{AV} = 80(\%) \cdot 0.06 = 4.8\,\%$$

**APPENDIX B**

The first threshold $th_{AB}(p)$ will be obtained by imposing the condition that the number of votes for the negative proposals of the majority party $P_1$ (due, as we know, to the contribution of $P_1$ itself and of 12,5% of independent legislators) is equal to $N/2 + 1$, i.e.:

$$(N - N_{ind})\frac{p}{100} + \frac{12.5}{100} N_{ind} = \frac{N}{2} + 1$$

Solving with respect to $N_{ind}$, we have:

$$N_{ind} = \frac{N(p-50) - 100}{(p-12.5)}$$

Then, dividing by $N$ and multiplying by 100, we obtain the threshold value in percentage:

$$th_{AB} = \frac{N(p-50)-100}{(p-12.5)} \frac{100}{N}$$

For *N*=500 and *p*=60 we have, as expected, $th_{AB}$=21% (corresponding to $N_{ind}$=103).

In a similar way, the second threshold $th_{BC}(p)$ will be obtained by imposing the condition that the number of votes for the positive proposals of the relative minority party $P_2$ (due to the contribution of all members of $P_2$, one half of $P_1$ and of 37.5% of independents) is equal to *N*/2 + 1, i.e.:

$$(N - N_{ind})\frac{100-p}{100} + (N - N_{ind})\frac{p}{100}\frac{1}{2} + \frac{37.5}{100}N_{ind} = \frac{N}{2} + 1$$

Going on as in the previous case, after some algebra we obtain:

$$th_{BC} = \frac{N(100-p) - 200}{(125-p)}\frac{100}{N}$$

that, for *N*=500 and *p*=60, gives, as expected, $th_{BC}$=61% (corresponding to $N_{ind}$=305).

Finally, the third threshold $th_{CD}(p)$ can be obtained by imposing the condition that the number of votes for the positive proposals of the majority party $P_1$ (due to the contribution of all members of $P_1$, one half of $P_2$ and of 37.5% of independents) is equal to N/2 + 1, i.e.:

$$(N - N_{ind})\frac{p}{100} + (N - N_{ind})\frac{100-p}{100}\frac{1}{2} + \frac{37.5}{100}N_{ind} = \frac{N}{2} + 1$$

Going on as in the two previous cases, we obtain:

$$th_{CD} = \frac{Np - 200}{(p+25)}\frac{100}{N}$$

that, for *N*=500 and *p*=60, gives, as expected, $th_{BC}$=70% (corresponding to $N_{ind}$=350).

**APPENDIX C**

Let us first derive, for each one of the four regions A, B, C and D addressed in section 4, the percentage of accepted proposals $N_{\%ACC}$ adds a function of the number of independent legislators $N_{ind}$:

A. In this region, $N_{\%ACC}$ is the sum of three terms, one due to the contribution of the positive proposals of the relative majority party $P_1$, another one due to the contribution of the positive proposals of the opposition party $P_2$, and a third one due to the contribution of the negative proposals of $P_1$:

$$N_{\%ACC-A}(N_{ind}) = p\frac{N-N_{ind}}{N}\int_0^1 P_1(Y)dY + (100-p)\frac{N-N_{ind}}{N}\int_0^1 P_2(Y)dY + p\frac{N-N_{ind}}{N}\int_{-1}^0 P_1(Y)dY \quad (1)$$

B. In this region $N_{\%ACC}$ is the sum of two terms, one due to the contribution of the positive proposals of P$_1$ and the other one due to the contribution of the positive proposals of P$_2$:

$$N_{\%ACC-B}(N_{ind}) = p\frac{N-N_{ind}}{N}\int_0^1 P_1(Y)dY + (100-p)\frac{N-N_{ind}}{N}\int_0^1 P_2(Y)dY \qquad (2)$$

C. In this region $N_{\%ACC}$ is due only to the contribution of the positive proposals of P$_1$:

$$N_{\%ACC-C}(N_{ind}) = p\frac{N-N_{ind}}{N}\int_0^1 P_1(Y)dY \qquad (3)$$

D. Finally, in this region there are no more contributions and $N_{\%ACC}$ is zero:

$$N_{\%ACC-D}(N_{ind}) = 0$$

(4)

What should we now expect, in the limit of many legislative terms, for the theoretical average value of the social gain $Y_{AV}$ produced by the accepted proposals inside each of the four regions? Of course, this strictly depends on the kind of proposals accepted:

A. In this interval, positive and negative accepted proposals coming from the relative majority party P$_1$ give, on average, a globally null value of the social gain; the only positive contribution to the social gain is given by the positive accepted proposals coming from the relative minority party P$_2$. But now we need, again, to distinguish between two different perspectives: that one of the parties, for which the average social gain of the positive proposals is worked out considering the parties themselves centered at Y=0, and that of the independent legislators, for which the average social gain of the positive proposals coming from the parties is centered at Y=0.5. This gives rise to two different contributions, weighted by the relative number of both parties and independent legislators:

$$Y_{AV-A}(N_{ind}) = \frac{N-N_{ind}}{N}\left[\int_{-1}^1 P_1(Y)Y\,dY + \int_0^1 P_2(Y)Y\,dY\right] + \frac{N_{ind}}{N}0.5 \qquad (5)$$

B. In this interval, the negative proposals of P$_1$ are no longer accepted, therefore the average social gain results from the contribution of the positive proposals only, coming from both P$_1$ and P$_2$. Again, the different perspectives of parties and independents give raise to two different contributions, weighted by their relative number:

$$Y_{AV-B}(N_{ind}) = \frac{N-N_{ind}}{N}\left[\int_0^1 P_1(Y)Y\,dY + \int_0^1 P_2(Y)Y\,dY\right] + \frac{N_{ind}}{N}0.5 \qquad (6)$$

C. In this interval, the average social gain is only due to the contribution of positive proposals of P$_1$, where – again – one has to distinguish the two terms representing, respectively, the parties' perspective and the independent legislators' perspective:

$$Y_{AV-C}(N_{ind}) = \frac{N-N_{ind}}{N}\left[\int_0^1 P_1(Y)Y\,dY\right] + \frac{N_{ind}}{N}0.5 \qquad (7)$$

D. In this interval no more proposal are accepted at all, therefore the average social gain should be null:

$$Y_{AV-D}(N_{ind}) = 0 \qquad (8)$$

We expect that this it is strictly true only in the limits of an infinite number of both legislators and legislative terms: for finite numbers of them, fluctuations in the Gaussian distributions of parties and in their positions on the Y axis will make the null prediction an underestimation of the numerical results.

**APPENDIX D**

Exploiting the results of Appendix C, we calculate now the expected average global efficiency $Eff_{exp}$ of our parliament as a function of p, N and $N_{ind}$, in the four regions A, B, C and D.

To do so, we preliminarily substitute in all the obtained equations the various definite integrals with their effective values, which do not depend on p, N and $N_{ind}$ but are fixed. These values, which follow from the normalization of the probability distributions of the proposals, are:

(1) $\int_{-1}^{1} P_i(Y) dY = 1 \; ; \quad \int_{0}^{1} P_i(Y) dY = 0.5 \; ; \quad \int_{-1}^{1} P_i(Y) Y \, dY = 0 \; ; \quad \int_{0}^{1} P_i(Y) Y \, dY = 0.06 \quad for \; i = 1,2$

Let us start with region A (with $0 < N_{ind} < th_{AB}$), where one has:

$$Eff_A(p, N, N_{ind}) = N_{\%ACC-A} \cdot Y_{AV-A} = \left( p \frac{N - N_{ind}}{N} + (100 - p) \frac{N - N_{ind}}{N} 0.5 \right) \cdot \left( \frac{N - N_{ind}}{N} 0.06 + \frac{N_{ind}}{N} 0.5 \right)$$

from which, after some algebra:

(2) $$Eff_A(p, N, N_{ind}) = (1 - \frac{N_{ind}}{N})(50 + \frac{p}{2})(0.06 + 0.44 \frac{N_{ind}}{N})$$

The derivative which (for any p) is always positive within the interval considered:

$$\frac{d}{dx} Eff_A(p,x) = \frac{d}{dx}(1-x)(50+\frac{p}{2})(0.06+0.44x) = p(0.19-0.44x) - 44x + 19 > 0 \quad for \; x < 0.43$$

Let's now continue with region B (with $th_{AB} < N_{ind} < th_{BC}$), where we find:

$$Eff_B(p, N, N_{ind}) = N_{\%ACC-B} \cdot Y_{AV-B} = \left( p \frac{N - N_{ind}}{N} 0.5 + (100 - p) \frac{N - N_{ind}}{N} 0.5 \right) \cdot \left( \frac{N - N_{ind}}{N}(0.06 + 0.06) + \frac{N_{ind}}{N} 0.5 \right)$$

from which one obtains:

(3) $$Eff_B(N, N_{ind}) = 50 \, (1 - \frac{N_{ind}}{N})(0.12 + 0.38 \frac{N_{ind}}{N})$$

Notice that inside this region $Eff_B$ does not depend on $p$. Furthermore, in its maximum possible range (i.e. between 0 and $th_{BC}(max)=65\%$), it has always a maximum whose value can be obtained by equating to zero the derivative:

$$\frac{d}{dx}Eff_B(p,x) = \frac{d}{dx}50\,(1-x)(0.12+0.38x) = 13-38x = 0 \;\;\rightarrow\;\; x = \frac{13}{38} \approx 0.34$$

For the region C (with $th_{BC} < N_{ind} < th_{CD}$), one has:

$$Eff_C(p,N,N_{ind}) = N_{\%ACC-C} \cdot Y_{AV-C} = \left(p\frac{N-N_{ind}}{N}0.5\right)\cdot\left(\frac{N-N_{ind}}{N}0.06 + \frac{N_{ind}}{N}0.5\right)$$

from which the following expression can be derived:

$$Eff_C(p,N,N_{ind}) = (1-\frac{N_{ind}}{N})\frac{p}{2}\left(0.06 + 0.44\frac{N_{ind}}{N}\right) \tag{4}$$

The derivative (for any p) is always negative in the interval considered:

$$\frac{d}{dx}Eff_C(p,x) = \frac{d}{dx}(1-x)\frac{p}{2}(0.06+0.44x) = p(0.19-0.44x) < 0 \;\;\; for\;\; x > 0.43$$

Finally, for region D (with $th_{CD} < N_{ind} < N$):

$$Eff_D(p,N,N_{ind}) = N_{\%ACC-D} \cdot Y_{AV-D} = 0 \tag{5}$$

# References

Aristotle. *The Athenian Constitution,* trans. Moore, J. M. In *Aristotle and Xenophon on Democracy and Oligarchy,* Berkeley: University of California Press, 1986.

Arrow, K.J. *Social Choice and Individual Values*, 2nd ed., New York: John Wiley & Sons, 1963.

Bowler, Shaun, David M. Farrell, and Richard S. Katz eds. *Party Discipline and Parliamentary Government*, Columbus: The Ohio State University Press, 1999.

Callenbach, Ernest., Michael Phillips, and Keith Sutherland. *A Citizen Legislature*, Exeter: Imprint Academic, 2008 [originally published in 1985 by Banyan Tree Books/Clear Glass].

Chan, William, and Priscilla Man. "Help and Factionalism in Politics and Organizations", *Southern Economic Journal* 79.1 (2012): 144-160.

Delannoi, Gil. and Oliver Dowlen and Peter Stone. *The Lottery As a Democratic Institution*, Dublin: Trinity College, 2013.

Delannoi, Gil, and Oliver Dowlen. *Sortition: Theory and Practice*, Vol. 3, Luton: Andrews UK Limited, 2016.

De Tocqueville, Alexis. *De la Démocratie en Amérique*. Vol. 1, Paris: Pagnerre, 1850.

Dowlen, Oliver. "Sorting out Sortition: A Perspective on the Random Selection of Political Officers", *Political Studies* 57.2 (2009): 298-315.

Dowlen, Oliver. *The Political Potential of Sortition: A Study of the Random Selection of Citizens for Public Office*, Vol. 4, Luton: Andrews UK Limited, 2017.

Eguia, Jon X. "Voting blocs, party discipline and party formation", *Games and Economic Behavior* 73.1 (2011): 111-135.

Frey, Bruno. "Proposals for a Democracy of the Future", *Homo Oeconomicus, 34*.1 (2017): 1-9.

Köppl-Turyna, Monika. "Some Thoughts on Frey's 'Proposals for a Democracy of the Future'", *Homo Oeconomicus* 35.1 (2018): 1-5.

Lockard, Alan A. "Decision by Sortition: a Means to Reduce Rent-Seeking", *Public Choice*, 116.3 (2003): 435-451.

Najemy, John M. *Corporatism and Consensus in Florentine Electoral Politics, 1280-1400*, Chapel Hill: University of North Carolina Press, 1982.

Pluchino, Alessandro, Andrea Rapisarda and Cesare Garofalo. "The Peter Principle Revisited: a Computational Study" *Physica A: Statistical Mechanics and Its Applications*, 389.3 (2010): 467-472.

Pluchino, Alessandro, Cesare Garofalo, Andrea Rapisarda, Salvatore Spagano and Maurizio Caserta. "Accidental Politicians: How Randomly Selected Legislators can Improve Parliament Efficiency", *Physica A: Statistical Mechanics and Its Applications*, 390.21-22 (2011): 3944-3954.

Statuta Communis Parmae, *in* "Monumenta Historica ad provincias Parmensem et Placentinam pertinentia", Parma (1855).

Stone, Peter. *The Luck of the Draw: The Role of Lotteries in Decision Making*, Oxford: Oxford University Press, 2011.

Stone, Peter. "Sortition, Voting, and Democratic Equality", *Critical Review of International Social and Political Philosophy* 19.3 (2016): 339-356.

Tridimas, George. "On sortition. Comment on 'Proposals for a Democracy of the Future' by Bruno Frey", *Homo Oeconomicus* 35.1-2 (2018): 91-100.

Villani, Giovanni. *Cronica di Giovanni Villani*, Ignazio Moutier and Francesco Gherardi Dragomani eds., Florence, 1845.

Wolfson, Arthur M. "The Ballot and other Forms of Voting in the Italian Communes", *The American Historical Review* 5.1 (1899): 1-21.